\documentclass[aps,prl,twocolumn,superscriptaddress]{revtex4}
\usepackage{bm}
\usepackage{amsmath}
\usepackage{amssymb}
\usepackage[dvips]{graphicx}

\begin{document}

\title{ Hydrodynamic Self-Consistent Field Theory \\
	for Inhomogeneous Polymer Melts }

\author{David M. Hall}
\affiliation
        {
            Department of Physics,
            University of California Santa Barbara,
            Santa Barbara, CA 93106
        }

\affiliation
        {
            Los Alamos National Laboratory,
            Theoretical  Division,
            Los Alamos, NM 87545
        }

\author{Turab Lookman}
\affiliation
        {
            Los Alamos National Laboratory,
            Theoretical  Division,
            Los Alamos, NM 87545
        }
\author{Glenn H. Fredrickson}
\affiliation
        {
            Materials Research Laboratory,
            University of California Santa Barbara,
            Santa Barbara, CA 93106
        }
\affiliation
        {
            Department of Chemical Engineering,
            University of California Santa Barbara,
            Santa Barbara, CA 93106
        }
\author{Sanjoy Banerjee}
\affiliation
        {
            Department of Chemical Engineering,
            University of California Santa Barbara,
            Santa Barbara, CA 93106
        }
\date{\today}

\begin{abstract}

We introduce a mesoscale technique for simulating the structure and rheology
of block copolymer melts and blends in hydrodynamic flows. The technique
couples dynamic self consistent field theory (DSCFT) with continuum hydrodynamics
and flow penalization to simulate polymeric fluid flows in
channels of arbitrary geometry. We demonstrate the method by studying phase separation of
an ABC triblock copolymer melt in a sub-micron channel with neutral wall wetting conditions.
We find that surface wetting effects and shear effects compete, producing wall-perpendicular
lamellae in the absence of flow, and wall-parallel lamellae in cases where 
the shear rate exceeds some critical Weissenberg number.
  
\end{abstract}
\maketitle

Predicting the non-equilibrium dynamics of multi-component 
inhomogeneous polymer formulations is important for
development and improvement of paints, adhesives, cosmetics, 
and processed foods. Many such systems have fluid structures
with length scales spanning nanometers to millimeters, and
stabilization of the finest scale structures is often
assisted by the presence of block or graft copolymers
\cite{FredricksonBook}. Computational tools which predict 
equilibrium properties of polymeric materials
\cite{CombinatorialScreening, FredricksonDrolet} 
have been developed as a cost effective alternative 
to trial and error experimentation. However, bulk properties
such as strength, optical clarity, and elasticity often depend in detail
on the size and distribution of inhomogeneities in the system,
which are determined both by the composition of the fluid 
\emph{and} the manner in which it was processed.

We have developed a mesoscale technique aimed at examining
the effects of hydrodynamic transport on polymer self assembly 
and viscoelasticity in industrial processing flows. The scheme couples
dynamic self consistent field theory (DSCFT)
\cite{FraaijeMesodyn,ReisterSpinodal, HasegawaDoi, KawakatsuODT} 
with a multiple-fluid Navier-Stokes
system and a set of constitutive equations which describe viscoelastic
stresses in a polymeric fluid. Both the thermodynamic and hydrodynamic portions 
of the model employ rigid wall fields which represent a slightly 
porous material that may be shaped to form channels with arbitrarily complex geometries. 
Each rigid object in the system may be fixed or moving, simulating the 
interaction of machine elements with polymers in the melt.
We refer to the scheme as hydrodynamic self consistent field theory (HSCFT) 
to underscore its emphasis on hydrodynamic transport. 

In contrast to ``phase field'' techniques \cite{ZhangDiblocks,ViscoelasticPhaseSeparation,GersappeShearedBlends},
this method simulates the assembly of multiblock copolymer mesophases where the 
polymeric nature of the chains is explicitly taken into account.
It also differs from SCFT and DSCFT techniques in its ability
to model hydrodynamic transport in complex channels with moving machinery.
Generally speaking, SCFT is capable of describing equilibrium morphologies 
and micro-phases boundaries. DSCFT is capable of describing non-equilibrium
systems in which hydrodynamic transport may be neglected, 
including phase separating melts and systems subjected to simple shear fields.
HSCFT, in contrast, is appropriate for the description of non-equilibrium systems in which
hydrodynamic effects play an important role including dynamic polymer nanocomposites,
microfluidic channel flows, and micro-scale rheometry of material properties.

In this letter, we introduce a multi-fluid generalization of the ``two-fluid" model
of Doi and Onuki \cite{TwoFluidModel} followed by a summary of the thermodynamic equations
used to model block-copolymer self assembly in the presence of rigid boundary surfaces.
We illustrate the method by examining the effects of hydrodynamic transport on a
phase separating triblock copolymer melt in a narrow channel. 

In the most general case, the system is composed of a blend 
of $C$ distinguishable copolymer species in a fixed volume, $V$.
Each copolymer of species $\alpha$ is itself constructed from $M_\alpha^\phi$
distinct monomer types (such as polystyrene, polyethylene, etc.)
Each copolymer species $\alpha$ is also taken to be monodisperse with
a polymerization index $N_\alpha$ and an entanglement length $N_{e\alpha}$. 
In addition, the system contains a set of rigid walls which are constructed from  
$M^\psi$ distinguishable solid materials. 

We derive a hydrodynamic model for multiple viscoelastic fluids at low Reynolds number
by employing the principles of irreversible thermodynamics as outlined in \cite{TwoFluidModel}. 
The method described therein consists of a formal procedure for constructing
a Rayleigh functional $\mathcal{R}=\dot{F}+(1/2)W$ where $\dot{F}$ represents
the time derivative of the free energy and $W$ is a dissipation functional. 
Extremizing $\mathcal{R}$ with respect to the component velocities produces
hydrodynamic equations of motion while preserving the correct Onsager couplings.
Due to space constraints, we present the results of the derivation here, 
deferring the details to a later publication.

The fluid composition at a given point $\mathbf{r}$ is described by monomer volume fraction fields
 $\phi_{\alpha i}(\mathbf{r})$ and wall volume fraction fields $\psi_j(\mathbf{r})$. 
The first index on  $\phi_{\alpha i}(\mathbf{r})$ indicates the copolymer species, 
and the second index indicates the material type (polystyrene, polyethylene, etc.)
Note that in general, polystyrene monomers associated with one copolymer species $\phi_{11}(\mathbf{r})$
exhibit different dynamics from polystyrene monomers associated with 
a second copolymer species $\phi_{21}({\mathbf{r}})$ requiring us to treat them as distinct fluids.
The field $\psi_j(\mathbf{r})$ indicates the volume fraction of solid material 
of type $j$ at location $\mathbf{r}$. By definition, all volume fraction fields sum to unity, 
$\Sigma_{\alpha i} \phi_{\alpha i} + \Sigma_{j} \psi_j =1$. To save space, we employe
 the shorthand notation 
$\Sigma_{\alpha i}=\Sigma_{\alpha=1}^C\Sigma_{i=1}^{M^\phi_\alpha}$ 
and $\Sigma_j = \Sigma_{j=1}^{M^\psi}$.

In the absence of chemical reactions, the number density of each monomer species is conserved, as expressed by the continuity equations
\begin{eqnarray}
    \partial_t{\phi}_{\alpha i} \!\!+ 
    \nabla \cdot \phi_{\alpha i}\mathbf{v}^\phi_{\alpha i} \label{eqn:liquidTransport}=0\\
    \partial_t{\psi}_j \! 
    + \nabla \cdot \psi_j \mathbf{v}^\psi_j=0  \label{eqn:solidTransport}
\end{eqnarray}
where $\mathbf{v}^{\phi}_{\alpha i}(\mathbf{r})$ is the velocity of fluid $\phi_{\alpha i}$ and  $\mathbf{v}^{\psi}_j(\mathbf{r})$ is the velocity of the solid material $\psi_j$.

Each velocity field may be split into two parts
$\mathbf{v}^\phi_{\alpha i} =\mathbf{v}_T + \mathbf{w}^\phi_{\alpha
i}$. The tube velocity $\mathbf{v}_T = \sum_{\alpha i} \alpha_{\alpha
i}^\phi \mathbf{v}_{\alpha i}^\phi + \sum_j \alpha_j^\psi
\mathbf{v}_j^\psi$ represents the motion of the network of topological 
constraints in an entangled polymer melt as described 
in Brochard's theory of mutual diffusion \cite{Brochard}, and
$\mathbf{w}^{\phi}_{\alpha i}$ represents the velocity of a given
fluid relative to $\mathbf{v}_T$. The stress division parameters $\alpha_{\alpha i}^\phi(\mathbf{r})$
and $\alpha_j^\psi(\mathbf{r})$ are obtained by balancing the frictional forces 
acting on the network as shown in \cite{TwoFluidModel}.

The relative velocity fields are given by 
\begin{equation}
 \mathbf{w}^\phi_{\alpha i}  =
    \frac{1}{\zeta^\phi_{\alpha i}} \!\left[
    \alpha^\phi_{\alpha i} \nabla \cdot \bm{\sigma}
     -\phi_{\alpha i}\nabla \mu^\phi_{\alpha i}
     - \phi_{\alpha i}\nabla p + \phi_{\alpha i}\mathbf{f}_e\right]
     \label{eqn:relativeFlow}
\end{equation}
which are produced as a result of imbalanced osmotic forces $\phi_{\alpha i} \nabla \mu_{\alpha i}$,
pressure gradients $\nabla p$, viscoelastic forces $\nabla \cdot \bm{\sigma}$,
and external body forces $\mathbf{f}_e$. 
The relative velocity of each fluid is inversely proprtional to its friction coefficient  
$\zeta_{\alpha i}^\phi = \frac{1}{v_0}\zeta^\phi_{0i} \phi_{\alpha i} N_\alpha/N_{e\alpha}$ 
where $\zeta^\phi_{0i}$ is the monomer friction coefficient of material $i$ 
and $v_0$ is the volume occupied by a single monomer.

The local force imposed on the fluid by each solid object is
\begin{equation}
 \mathbf{f}_j^\psi(\mathbf{r}) = \psi_j \nabla \mu_{j}^\psi
     + \psi_j\nabla p
     - \alpha^\psi_j \nabla \cdot \bm{\sigma}
     +\zeta^\psi_j(\mathbf{v}^\psi_j - \mathbf{v}_T) \label{eqn:solidForces}
\end{equation}
In the case where the wall velocities  $\mathbf{v}^\psi_j$ are specified,
we may integrate over $\mathbf{f}_j^\psi(\mathbf{r})$ to obtain the net force and net torque
on each object, potentially enabling numerical rheometric experiments.

In the limit of low Reynolds number, the inertial terms may be neglected, and the momentum transport equation becomes a force balance equation.
\begin{equation}
    0 = \nabla \pi + \nabla p - \nabla \cdot \bm{\sigma} - \mathbf{f}^\psi - \mathbf{f}^\phi
    \label{eqn:momentumBalance}
\end{equation}
This equation implicitly determines the mean velocity field
 $\mathbf{v}=\Sigma_{\alpha i} \phi_{\alpha i} \mathbf{v}^\phi_i + \Sigma_j \psi_j \mathbf{v}^\psi_j$
which balances the net osmotic force
$\nabla \pi = \Sigma_{\alpha i} \phi_{\alpha i} \nabla \mu^\phi_i + \Sigma_j\psi_j \nabla \mu^\psi_j$,
 the pressure gradient $\nabla p$, the viscoelastic forces $\nabla \cdot \bm{\sigma}$, the
net wall force $\mathbf{f}^\psi$, and the net body force acting on the fluid $\mathbf{f}^\phi$. 

Energy is dissipated in the system by each of the terms
in the dissipation functional $W$\cite{TwoFluidModel},
including friction due to the motion of each fluid relative to the polymer network 
$\zeta_{\alpha i}(\mathbf{v}_{\alpha i}-\mathbf{v}_T)$, 
friction between the fluid and the walls 
$\zeta^\psi_{j}(\mathbf{v}^\psi_{j}-\mathbf{v}_T)$, 
and elastic deformation of the polymer network $\bm{\sigma}:\nabla \mathbf{v}_T$.
Note that in the absence of viscoelastic effects, the quantity $\nabla \cdot \bm{\sigma}$
reduces to the usual viscous dissipation term $\eta \nabla^2v$.

The mean velocity $\mathbf{v}$ and pressure fields $p$ must simultaneously
satisfy the force balance condition Eq.~(\ref{eqn:momentumBalance})
and the mass conservation condition, $\nabla \cdot \mathbf{v}=0$. These
relationships are not easily inverted and are obtained using an
iterative numerical procedure.  Once  $\mathbf{v}$ and $\mathbf{w}^\phi_j$ 
are known, the tube velocity may be calculated using the relationship
\begin{equation}
    \mathbf{v}_T = \frac{\sum_{\alpha i} (\alpha_{\alpha i}^\phi-\phi_{\alpha i})\mathbf{w}_i^\phi
                + \sum_j (\alpha_j^\psi -\psi_j)\mathbf{v}_i^\psi + \mathbf{v}}
                {1-\sum_{\alpha i} (\alpha_{\alpha i}^\phi-\phi_{\alpha i})}
\end{equation}

Constitutive equations (\ref{ShearStressEv}) and
(\ref{BulkStressEv})  are required to measure the bulk elastic stresses
$\bm{\sigma}^b_{\alpha i}$ and shear elastic stresses
$\bm{\sigma}^s_{\alpha i}$ induced by deformations of each entangled
polymeric component.
\begin{eqnarray}
	 \overset{\triangledown}{\bm{\sigma}}{}^s_{\alpha i} &=& 
 	G_{\alpha i} \bm{\kappa}_T - \bm{\sigma}^s_{\alpha i}/\tau_\alpha \label{ShearStressEv} \\
 	\overset{\triangledown}{\bm{\sigma}}{}^b_{\alpha i} &=& 
	K_{\alpha i} (\nabla \cdot \mathbf{v}_T)\bm{\delta} -\bm{\sigma}^b_{\alpha i}/\tau_\alpha
 	\label{BulkStressEv}
\end{eqnarray}
Although there are many possible constitutive equations, we have
chosen to employ the Oldyroyd-B model in which the total
viscoelastic force is produced by the sum of the elastic forces
contributed by each polymeric component and a dissipative viscous
force,
 $\nabla \cdot \bm{\sigma}= \sum_{\alpha i}(\nabla \cdot
\bm{\sigma}^s_{\alpha i} +\nabla \cdot  \bm{\sigma}^b_{\alpha i}) +
\eta_s \nabla^2 \mathbf{v}_T$. 
This model allows us to study a purely viscous fluid
 ($G_{\alpha i}=K_{\alpha i}=0$), a purely viscoelastic melt
 ($\eta_s=0$), or anything in-between. The derivatives in 
 (\ref{ShearStressEv}) and (\ref{BulkStressEv}) are upper convected time derivatives, 
$ \overset{\triangledown}{\bm{\sigma}} =  \partial_t\bm{\sigma}
 + \mathbf{v}_T \cdot \nabla \bm{\sigma}
 - \bm{\sigma}\cdot \nabla \mathbf{v}_T
 - (\nabla \mathbf{v}_T)^\dagger \cdot  \bm{\sigma}$.
and the shear strain rate tensor is 
$ \bm{\kappa}^{ij}_T=\partial_i v_T^j+\partial_j v_T^i 
- \frac{2}{d}\left( \nabla \cdot \mathbf{v}_T \right)\delta_{ij}$.
The concentration dependent bulk moduli $K_{\alpha i}$
and shear moduli $G_{\alpha i}$ measure the stresses
imposed on each component by the applied strains. 
Following Tanaka's example \cite{ViscoelasticPhaseSeparation}, 
we employ moduli of the form 
$G_{\alpha i}(\phi) = G_{0 \alpha i} \phi_{\alpha i}^2$ and 
$K_{\alpha i}(\phi) = K_{0 \alpha i} \theta\left(\phi_{\alpha i} - f_{\alpha i} \right)$ 
where $\theta$ is the step function. 
The quantity $\tau_\alpha$ represents the reptative disentanglement time associated with
the characteristic entanglement length $N_e$ of copolymer $\alpha$.
For simulations with large elastic strains, this constitutive equation
may be replaced with a more sophisticated phenomenological model 
or one based upon the SCFT microphysics \cite{FredricksonRheology}.

A self consistent calculation produces the chemical potential gradients which
drive phase separation. The free energy of the melt is $F=U-TS$ 
where the enthalpy is given by
\begin{equation}
    U/kT=\!\frac{1}{v_0 \bar{N}}\! \!\int \!\!d\mathbf{r}  \!\left[
    \frac{1}{2}\sum_{\alpha i \beta j} \phi_{\alpha i}
    \bar{N}\chi^\phi_{\alpha i \beta j}  \phi_{\beta j}
    + \sum_{i\beta j}\psi_i \bar{N}\chi^\psi_{i \beta j}  \phi_{\beta j}\right]
\end{equation}
The $\chi^\phi$ and $\chi^\psi$ denote segment-segment and
segment-wall interactions, respectively, and the entropy is
\begin{equation}
    S/k  = \sum_\alpha n^\phi_\alpha \ln Q_\alpha
    +  \frac{1}{v_0 \bar{N}} \int d\mathbf{r}
    \sum_{\alpha i} \omega_{\alpha i}^\phi\phi_{\alpha i}
\end{equation}
The quantity $Q_\alpha$ is the partition function over all
configurations of a single copolymer species $\alpha$ in the
presence of the fields $\omega^\phi_{\beta i}$.

Dynamic SCFT methods postulate that the system variables may
be divided into two sets which relax on widely separated time scales.
A ``slow" set of parameters includes the elastic stresses and the conserved
fields, mass, momentum, and monomer number. The remaining
variables, including the conjugate chemical fields 
$\omega_{\alpha i}^{\phi}$ are ``fast" parameters, 
which satisfy the local thermodynamic equilibrium (LTE) conditions
$\tilde{\phi}_{\alpha i}[\omega] = \phi_{\alpha i}$ \cite{FredricksonRheology}. 
which are derived by extremizing the free energy with respect to the
conjugate fields $\omega^{\phi}_{\alpha i}(\mathbf{r})$.
The non-equilibrium volume fraction fields $\phi_{\alpha i}$ and $\psi_j$ 
give rise to imbalanced chemical potentials
$\mu^{\phi}_{\alpha i}(\mathbf{r}) \label{mu_phi}=\frac{\delta F}{\delta
\phi_{\alpha i}(\mathbf{r})} = \frac{kT}{\bar{N}v_0} \left[
\Sigma_{\beta j} \bar{N}\chi^\phi_{\alpha i \beta j} \phi_{\beta j} +
 \Sigma_k \bar{N} \chi^\psi_{k\alpha i} \psi_k - \omega^\phi_{\alpha i} \right]$  
 and
$\mu^{\psi}_i(\mathbf{r})  \label{mu_psi}=\frac{\delta F}{\delta \psi_i(\mathbf{r})} 
= \frac{kT}{\bar{N}v_0}  \Sigma_{\beta j} \bar{N}\chi^\psi_{i\beta j} \phi_{\beta j}$
which drive diffusion and phase separation.

\begin{figure}[htb] 
	a \hspace{.22\linewidth} b\hspace{.22\linewidth}  c\hspace{.22\linewidth} d \\
    \resizebox{\linewidth}{!}
    {
   	 \includegraphics{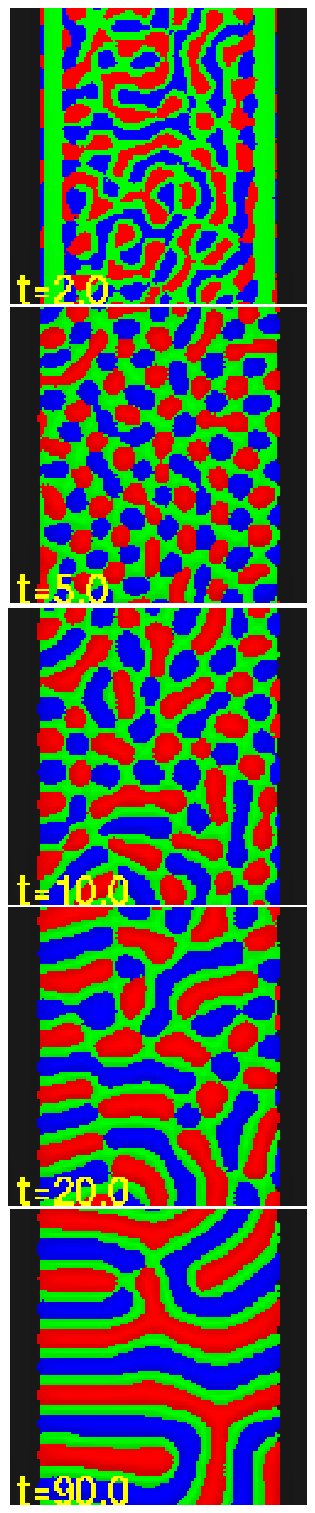}
  	  \includegraphics{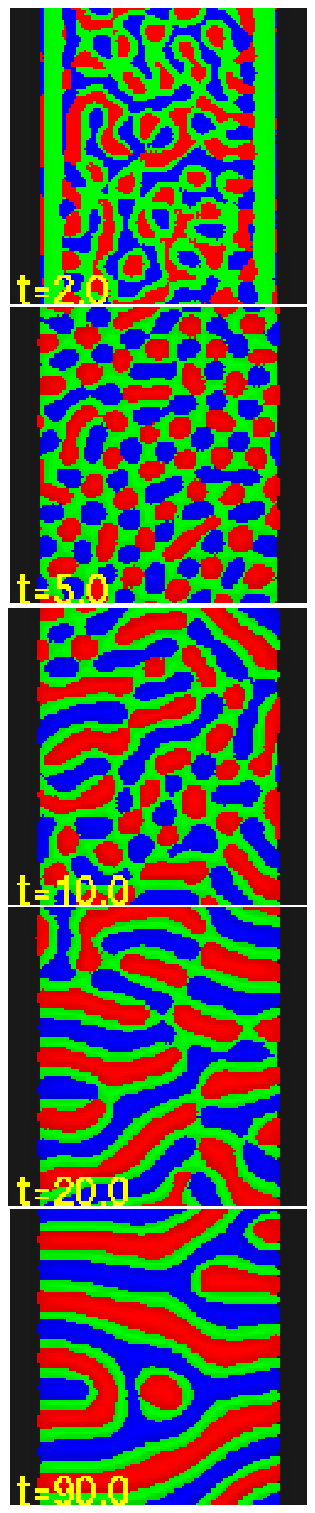}
  	  \includegraphics{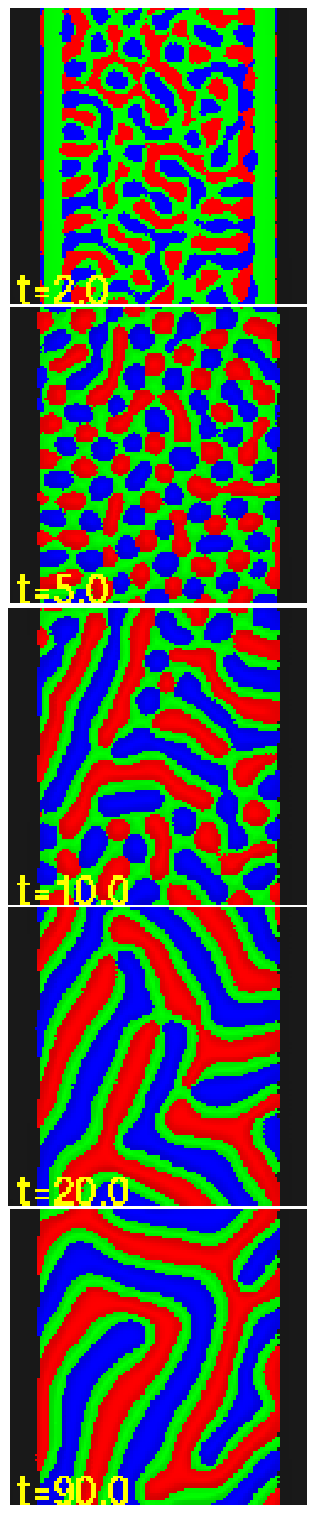}
   	 \includegraphics{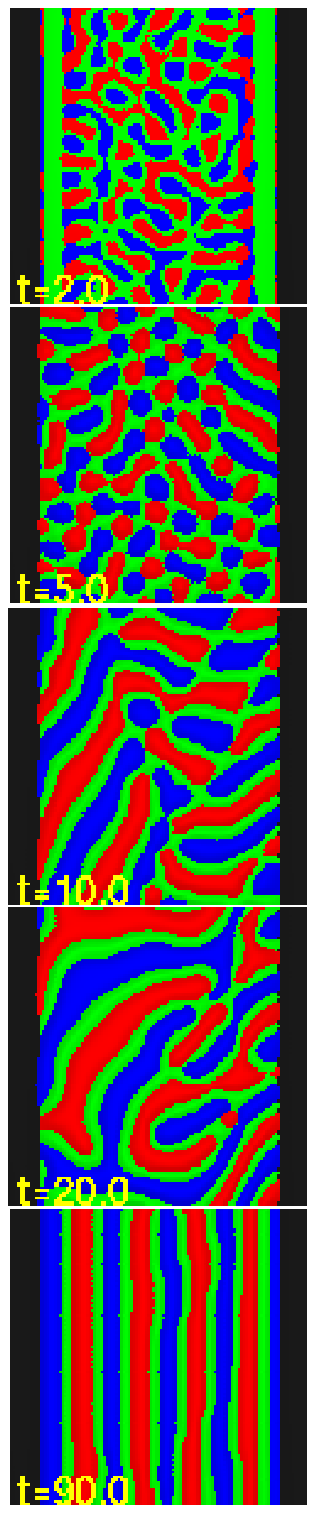}
    }
 
    \setlength{\unitlength}{\linewidth}
	\resizebox{\linewidth}{!}{\includegraphics{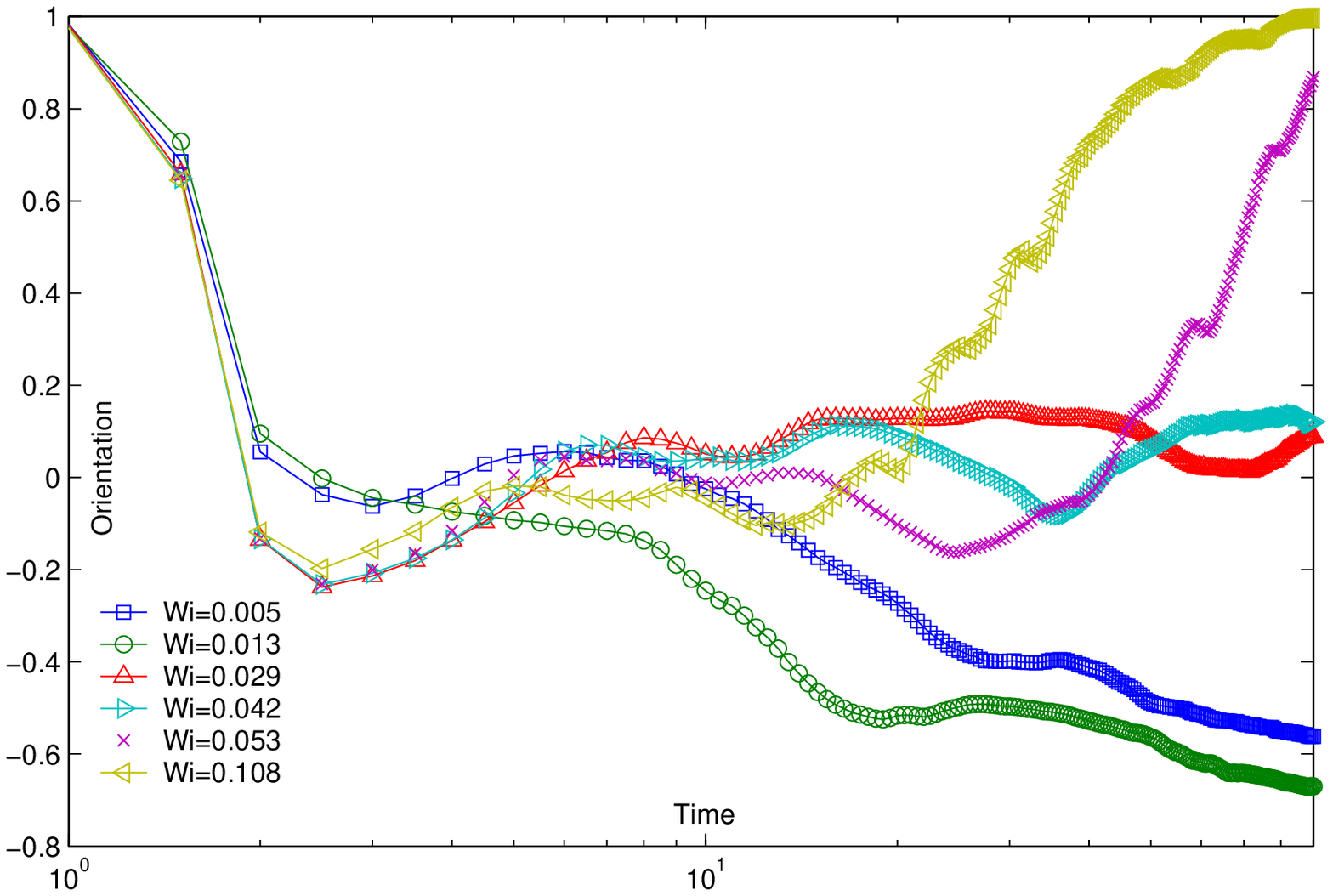}}
	\begin{picture}(0,0)
	\put(-0.3,+0.42){\resizebox{0.4\linewidth}{!}{\includegraphics{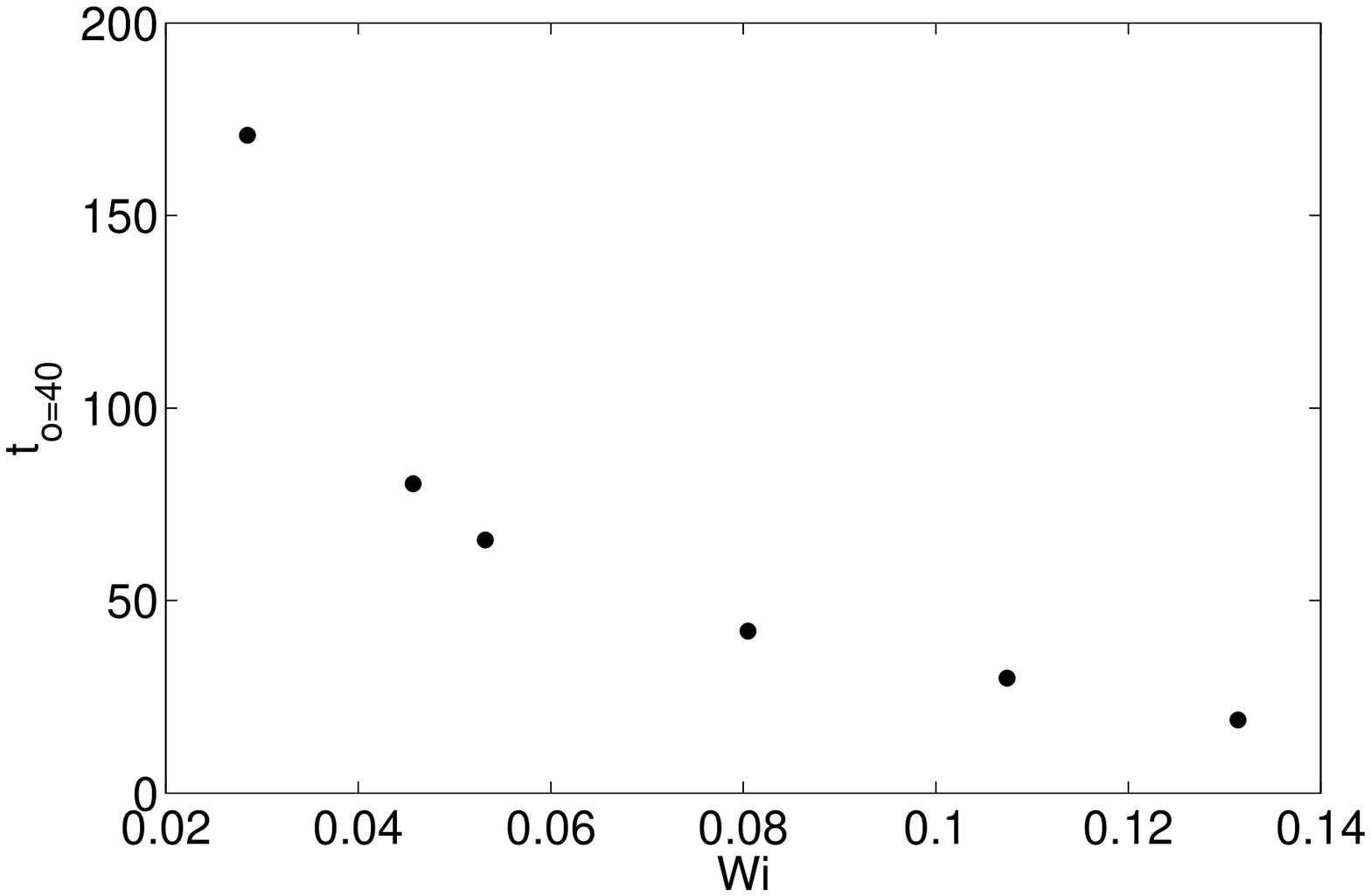}}}
	\end{picture}
  \caption{
  \label{FIG_MORPHOLOGY}
  [top] Time evolution of a quenched triblock melt in a sib-micron channel
   at several flow rates as characterized by the dimensionless Weissenberg number. 
    (a)  $\text{Wi}=0.005$ 
    (b)  $\text{Wi}=0.013$  
    (c)  $\text{Wi}=0.029$ 
    (d)  $\text{Wi}=0.108$ 
   [bottom] Lamellae orientation vs. time. The inset shows that time required to achieve $o_{xy}=0.40$ diverges as the Weissenberg number approaches $\text{Wi}=0.02$.}
\end{figure}

Calculations are performed in a periodic cell in two or three dimensions
in which non-periodic geometries may be constructed using rigid materials. 
Spatial derivatives are computed pseudo-spectrally in Fourier space 
which enables the resolution of sharp domain interfaces with a small number
of grid points and lends itself readily to parallelization. 
The multi-fluid model is solved iteratively for the
 $\bm{\sigma}, \mathbf{w}$, $\mathbf{f}^\psi$,$\mathbf{v}$ and $p$ fields.
Updated pressure and velocity fields are obtained
from a projection method and frictional forces penalize flow relative to the walls. 
The method repeats until a pseudo-steady state is achieved
which satisfies the no-flow, no-slip, continuuity, and force balance conditions (\ref{eqn:momentumBalance}). The new velocity fields are then used to transport the volume fraction fields.
Similarly, the local thermodynamic equilibrium conditions are solved by the 
iterative application of a hill climbing technique.

For concreteness, we have chosen to illustrate the technique 
by examining a single representative system in detail, 
namely the phase separation of a quenched triblock copolymer melt in a sub-micron channel.
The system consists of an incompressible melt of a symmetric ABC triblock copolymers ($f_i=1/3$)
quenched from a disordered state into the intermediate segregation regime,
in a channel of diameter $D=21R_g$. The Flory incompatibility parameters are fixed at
$\chi^\phi=25$, and $\chi^\psi=12$ and
all remaining parameters were chosen to approximate a polystyrene melt, with
$\zeta_0=10^{-8}\mbox{Ns/m}$ and $G_0=2000\mbox{kPa}$ \cite{frictionCoefficients}.
The flow rate is characterized by the dimensionless Weissenberg number
$\text{Wi} = \tau |\mathbf{v}_{\max}|/D$ where $\mathbf{v}_{\max}$ is the maximum mean velocity.


\begin{figure}[t] 
	a \hspace{.22\linewidth} b\hspace{.22\linewidth}  c\hspace{.22\linewidth} d \\
	 \resizebox{\linewidth}{!}
   	 {
	 \includegraphics[width=.5\linewidth]{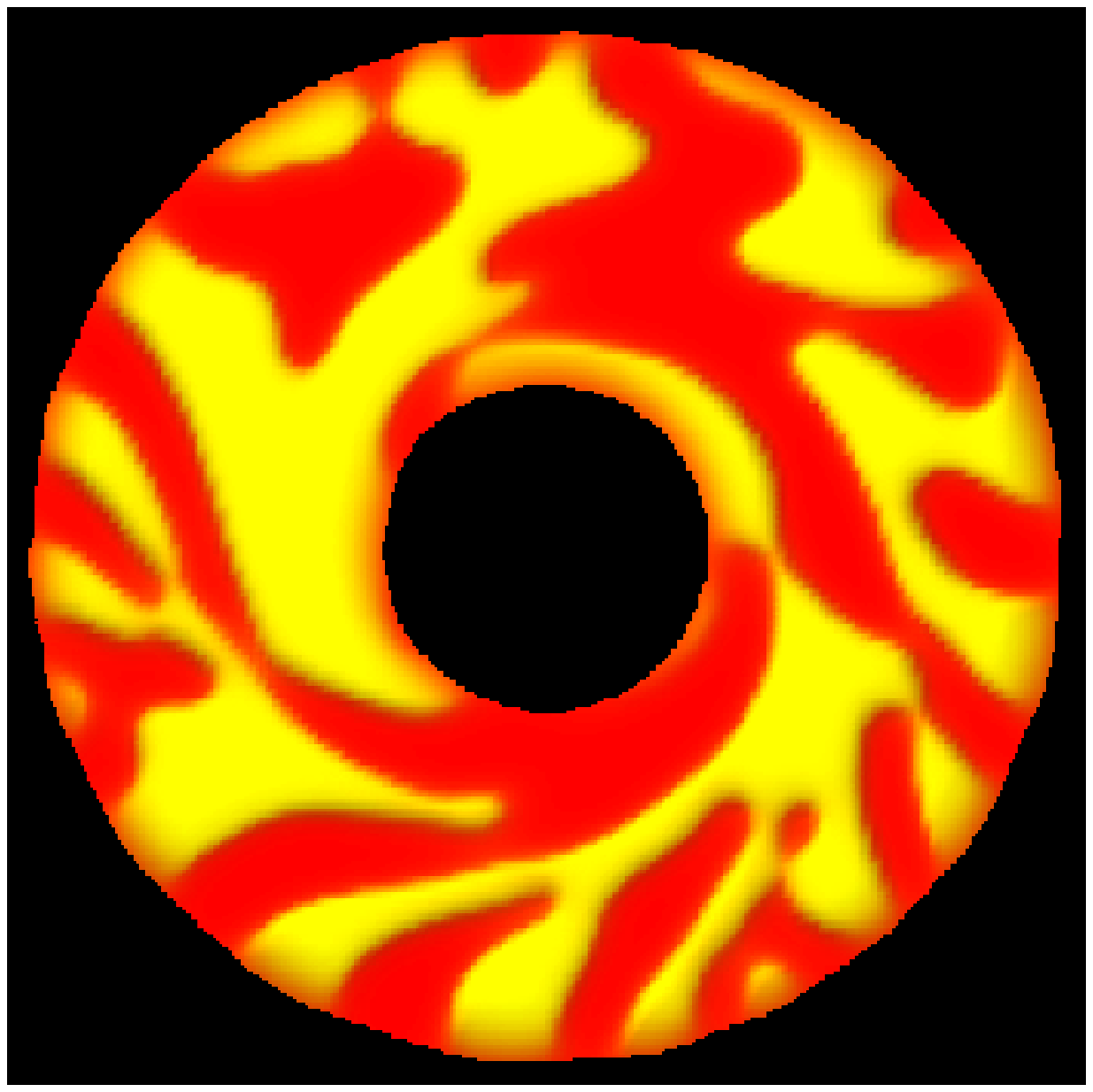}
	 \includegraphics[width=.5\linewidth]{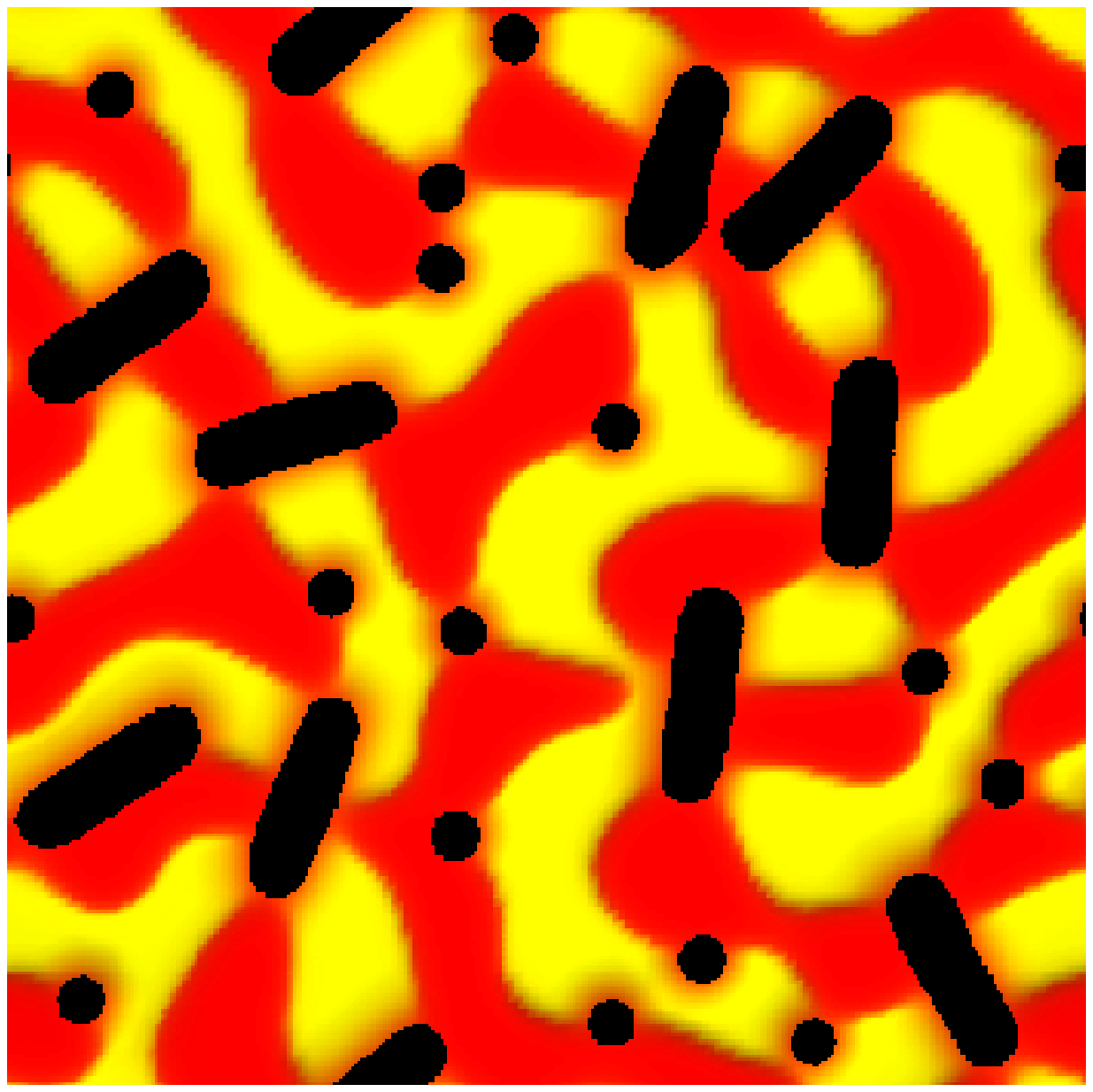}
    	 \includegraphics[width=.5\linewidth]{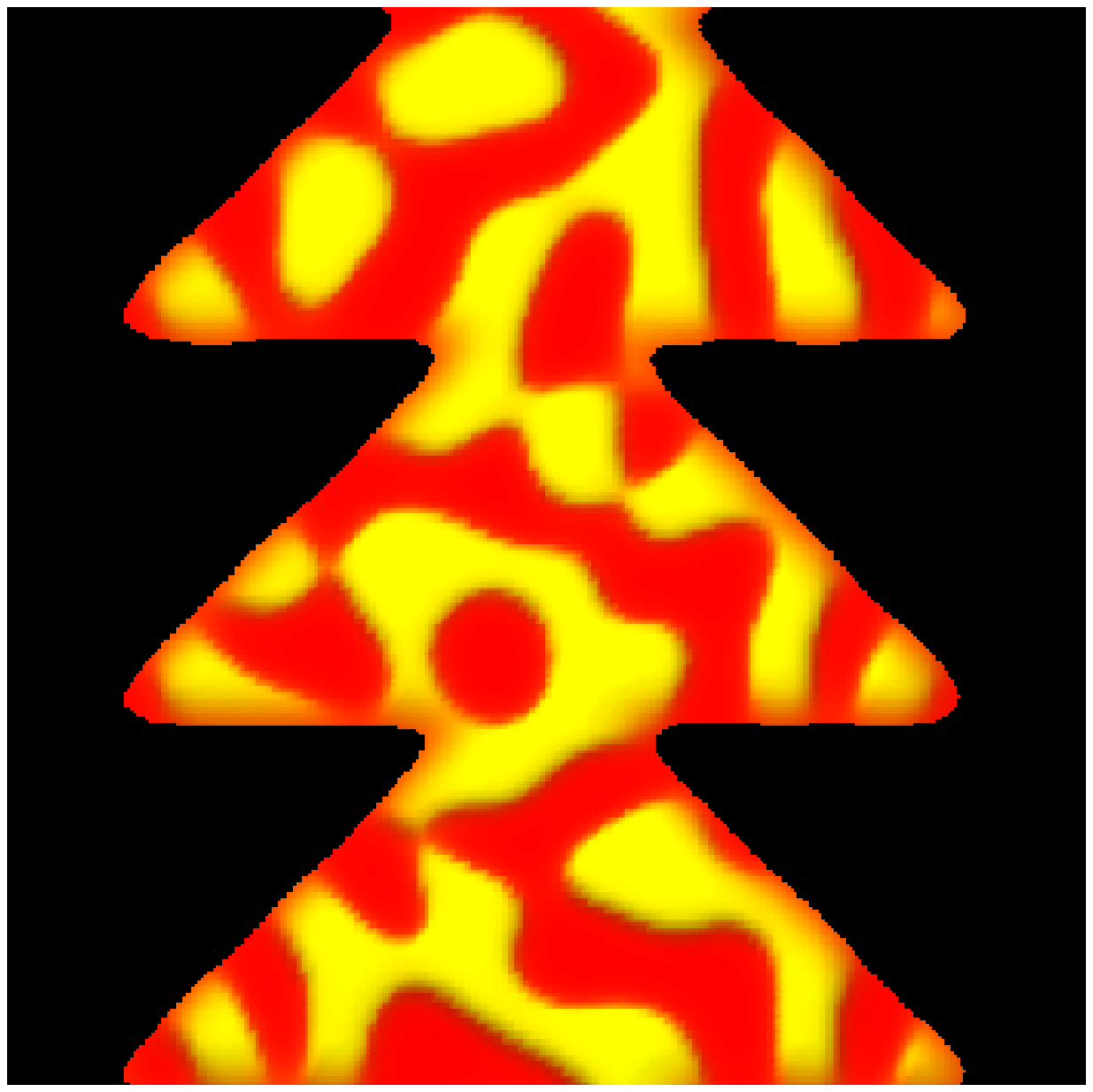}
    	 \includegraphics[width=.5\linewidth]{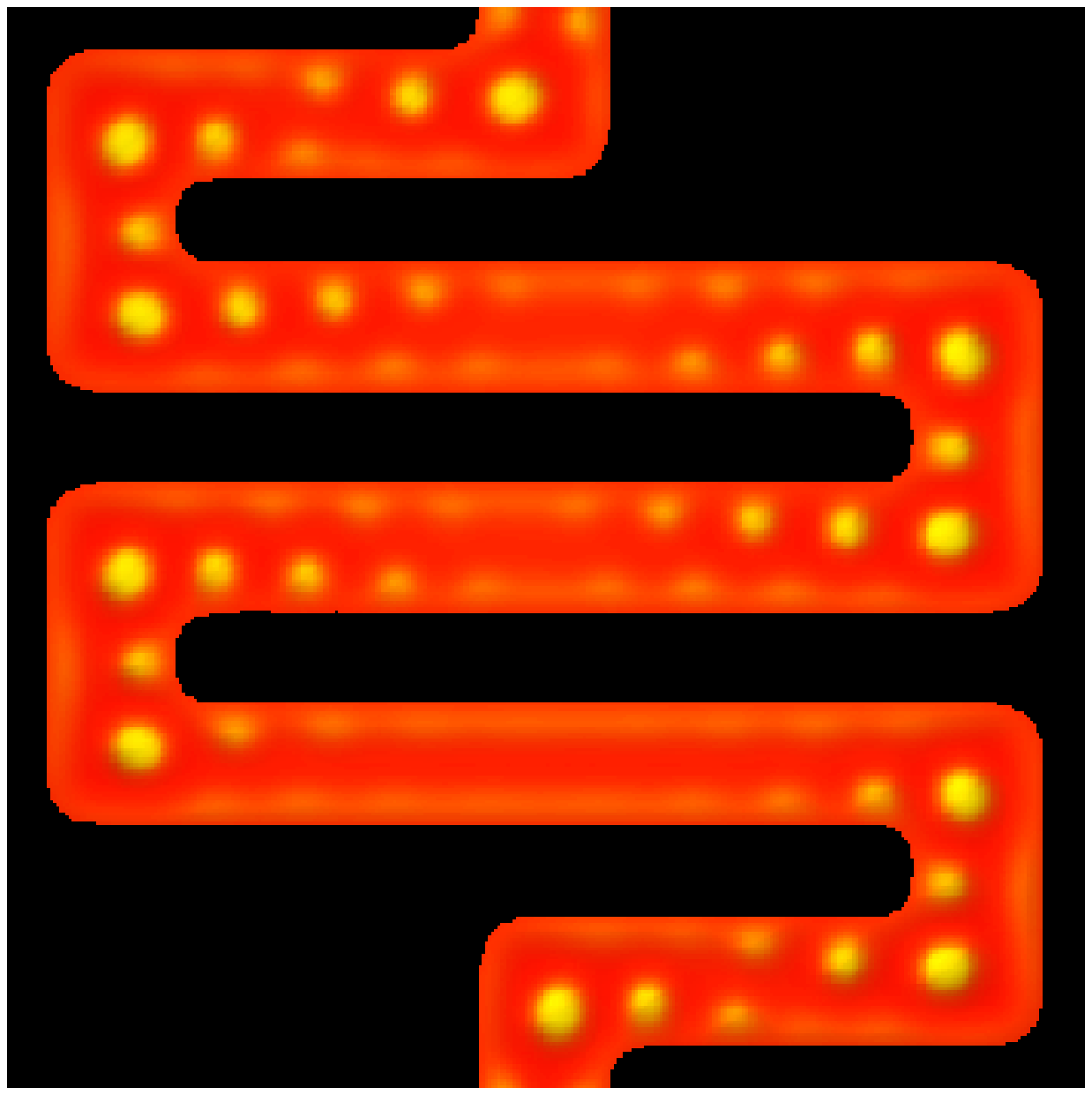}
	 }
	\caption{
	\label{FIG_EXAMPLES}
	Examples of 2D HSCFT systems: 
	(a) A homopolymer blend A+B in two concentric, rotating cylinders 
	(b) Nano-rods and nano-spheres in a phase separating blend 
	(c) Viscoelastic flow of a blend through a microfluidic rectifier 
	(d) An off critical diblock ($f=0.3$) in a micro-reactor channel.}
\end{figure}

The images in Fig.~\ref{FIG_MORPHOLOGY} illustrate the morphology 
as a function of time at various Weissenberg numbers, where
solid material is black and the copolymer blocks are red (A), green (B), and blue (C) respectively. 
We observe two distinct steady state configurations consisting of wall parallel 
and wall perpendicular lamellae, as illustrated by sequences (a) and (d) at $t=90$. 
In the absence of flow, boundary wetting conditions dominate, resulting in a perpendicular orientation, 
and at high flow rates shearing effects dominate, resulting in a parallel orientation with blocks arranged in a 
CBA ABC configuration. The two effects compete, and we anticipate that there is some minimum 
Wiessenberg number $\text{Wi}_c$ required to achieve a wall parallel configuration.

To quantify these observations, we construct an orientation parameter
$o_{xy}=\int \!d\theta s(\theta)[\cos^2\theta- \sin^2\theta] /(\int \!d\theta s(\theta))$
which measures the scattering difference between the x and y directions
in Fourier space, where $s(\theta)$ is the radially averaged scattering intensity.
The orientation parameter is plotted as a function of time
for several values of Wi in Fig.~\ref{FIG_MORPHOLOGY}, 
where $o_{xy}=1$ corresponds to wall-parallel lamellae
and $o_{xy}=-1$ corresponds to the wall-perpendicular conformation.

We estimate the value of the critical Weissenberg number, $\text{Wi}_c$, by plotting the time needed to achieve a parallel orientation, defined by $o_{xy}(t)>0.40$, 
as illustrated by the inset in Fig.~\ref{FIG_MORPHOLOGY}. 
It is clear from the figure that the alignment time diverges near the critical Weissenberg number,
allowing us to estimate $\text{Wi}_c\approx 0.02$ in a channel of diameter $D=21R_g$.
Furthermore, by comparing the two slowest cases, $\text{Wi}=0.005$ and $\text{Wi}=0.013$,
we observe that a flowing melt with $\text{Wi}\ll \text{Wi}_c$ approaches its steady state
configuration more rapidly than a quiescent melt. 
Similarly, by comparing the $\text{Wi}=0.053$ and $\text{Wi}=0.108$ cases,
we observe that a higher shear rate produces more rapid alignment in the $\text{Wi}\gg\text{Wi}_c$ regime.

This numerical experiment also illustrates a type of dynamic asymmetry which arises
in triblock copolymer melts. Although the system is 
compositionally symmetric, $f_A=f_B=f_C$, we see distinctly asymmetric structures 
in the early stages of phase separation as illustrated in Fig~\ref{FIG_MORPHOLOGY}. At $t\approx1$, 
we find wall parallel structures produced by preferential wetting of the walls by the A and C blocks, and
at $t\approx 5$, we see an asymmetric drop-in-matrix morphology, indicating that early
growth of density fluctuations is suppressed in regions which are enriched in block $B$ segments. 
Since the enthalpic terms are all identical, we conclude that these effects are produced solely due to
differences in the conformational entropy between the copolymer blocks.
To our knowledge this type of dynamic asymmetry has not previously been discussed.

In addition to the example provided above, HSCFT is capable of modeling systems of far greater complexity. As illustrated in Fig.~\ref{FIG_EXAMPLES}, some of the method's more unique capabilities include simulation of (a) the effects of moving machine parts on polymeric fluids (b) the unconstrained motion of solid nanoparticles (c) viscoelastic flow through irregular geometries and (d) flow in microfluidic channels.



In summary, we have developed a mesoscale technique which models
the effects of viscoelastic, Navier-Stokes type hydrodynamic transport 
on inhomogeneous polymeric fluids in arbitrarily complex channels.
In contrast with DSCFT methods, HSCFT 
is capable of describing the influence of moving machine elements,
nontrivial pressure induced and drag induced velocity fields,
and the unconstrained motion of solid nano-particles.
We believe this method has the potential for broad application and
represents a first step toward more realistic simulations of complex fluid flows in
industrial processing applications.

\bibliography{PREPRINT_PRL_LP10149_REV2}
\end{document}